\documentclass[]{spie}  
\usepackage[]{graphicx}
\usepackage{amsmath}
\usepackage{cite}

\title{Single photon emission and detection at the nanoscale utilizing semiconductor nanowires}

\author{Michael~E.~Reimer\supit{a}, Maarten~P.~van~Kouwen\supit{a}, Maria~Barkelid\supit{a}, Mo\"{i}ra~Hocevar\supit{a}, Maarten.~H.~M.~van~Weert\supit{a}, Rienk~E.~Algra\supit{b}, Erik~P.~A.~M.~Bakkers\supit{b}, Mikael~T.~Bj\"{o}rk\supit{c}, Heinz~Schmid\supit{c}, Heike~Riel\supit{c}, Leo~P.~Kouwenhoven\supit{a}, and Val~Zwiller\supit{a}
\skiplinehalf
\supit{a}Kavli Institute of Nanoscience, Delft University of Technology, Delft, The
Netherlands; \\
\supit{b}Philips Research Laboratories, Eindhoven, The Netherlands; \\
\supit{c}IBM Research GmbH, Zurich Research Laboratory, R\"{u}schlikon, Switzerland
}

\authorinfo{Further author information: (Send correspondence to V.Z.)\\M.E.R.: E-mail: m.e.reimer@tudelft.nl, Telephone: +31(0)15 278 6324 \\  V.Z.: E-mail: v.zwiller@tudelft.nl, Telephone: +31(0)15 278 6136}

\begin{document}
  \maketitle

\begin{abstract}
We report recent progress toward on-chip single photon emission and detection in the near infrared utilizing semiconductor nanowires. Our single photon emitter is based on a single InAsP quantum dot embedded in a p-n junction defined along the growth axis of an InP nanowire. Under forward bias, light is emitted from the single quantum dot by electrical injection of electrons and holes. The optical quality of the quantum dot emission is shown to improve when surrounding the dot material by a small intrinsic section of InP. Finally, we report large multiplication factors in excess of 1000 from a single Si nanowire avalanche photodiode comprised of p-doped, intrinsic, and n-doped sections. The large multiplication factor obtained from a single Si nanowire opens up the possibility to detect a single photon at the nanoscale.
\end{abstract}


\keywords{nanowires, single photon detection, quantum dot, electroluminescence, photoluminescence}

\section{INTRODUCTION}
\label{sec:intro}  

\indent
Semiconductor nanowires are particularly attractive in a variety of applications such as quantum information,\cite{vanWeert09, Kouwen10} LEDs,\cite{Duan01} solar cells,\cite{Tian07} and the detection of viruses\cite{Cui01} and gases\cite{Li04}. In terms of scalability, single nanowires can be controllably positioned,\cite{Pierret10} electrically contacted\cite{vanWeert09c, Kouwen10, vanWeert10} and integrated with Si-based technology\cite{Bakkers04}. In addition, the chemical doping along the nanowire can be controlled, thus allowing p-n and p-i-n structures to be fabricated \cite{vanWeert09c, Algra08} for light emission\cite{Minot07} and detection\cite{Yang06}. The one-dimensional device geometry of the nanowire has the added benefit that all of the electrically injected or extracted carriers are anticipated to flow through the active region of the device such as a single quantum dot. This feature is expected to increase the efficiency for converting an electrical signal into light or vice-versa\cite{Kouwen10APL} necessary for light generation and detection. Combined with plasmon waveguides,\cite{Heeres10}
these electro-optic devices potentially form the basis to perform
quantum optics experiments at the nanoscale, without the requirement
for a complex optical setup.

In the work presented here, we investigate single semiconductor
nanowires toward on-chip single photon emission and detection with the
aim of performing these tasks electrically in the near infrared. Similar to previous work, the single photon emitter is based on a single quantum dot, which has been shown to emit single photons\cite{Shields07, Claudon10} or
entangled photons\cite{Salter10} on demand. However, until now, most of these devices are based on self-assembled quantum
dots that are surrounded by a three-dimensional host matrix. In such
a device, if driven electrically, only a fraction of the injected
current is converted into an emitted photon and is therefore very
inefficient. It would be desirable to build a device in which 100\,$\%$
of the injected current would recombine from only a single
quantum dot. Our approach to achieve an efficient electron-photon
conversion is to embed a single InAsP quantum dot within the
depletion region of a p-n junction synthesized in an InP nanowire.
Although we have shown electrical injection into a single quantum
dot in previous work,\cite{Minot07} we show here that the dot
emission can be improved by surrounding the quantum dot with a small
intrinsic region of InP. The benefits of embedding single quantum
dots in nanowires to increase the efficiency of extracted light has
become apparent recently in the work of J. Claudon and co-workers.\cite{Claudon10}
In that work, an efficiency of 72 percent has been achieved for the
single photon source by embedding a single quantum dot in a tapered
nanowire.\cite{Claudon10} Their tapered nanowires have been
defined with top-down processing techniques, which may affect
out-coupling of light due to the surface roughness of the waveguide
caused by etching. Moreover, such a device has only been operated by
optical pumping with an external laser. Here, we utilize bottom-up fabrication approaches to synthesize nanowires and rely on electrical injection of carriers only. Finally, we present a single Si nanowire avalanche photodiode, which is comprised of p-doped, intrinsic and n-doped regions. We demonstrate here that multiplication factors in excess of 1000 are achieved from only a single Si nanowire. This multiplication factor is two orders of magnitude larger as compared to previously reported single
nanowire devices.\cite{Yang06} The large multiplication factor that
we obtain opens up the possibility to detect a single photon from
only a single nanowire device.

This paper is organized as follows: in Section \ref{sec:QLED} we first present our quantum light emitting diode (QLED), which is based on a single InAsP quantum dot embedded within the depletion region of a p-n junction synthesized in an InP nanowire. In Section \ref{sec:optical-properties}, the optical properties of single InAsP quantum dots in InP nanowires is introduced. Next, Section \ref{sec:p-n} shows that by simply applying a voltage across the nanowire comprised of a p-n junction, light is emitted by the single quantum dot. In Section \ref{sec:p-i-n}, the quantum dot is surrounded by a small intrinsic region of InP, which is shown to improve the optical properties of the quantum dot emission and potentially increase the quantum efficiency of the QLED. Finally, we present a single Si nanowire avalanche photodiode in Section \ref{sec:SiNW}. Details of the device are provided in Section \ref{sec:IV}. The relatively large multiplication factor obtained from only a single Si nanowire is presented in Section \ref{sec:Multiplication}. In Section \ref{sec:PC}, position dependent photocurrent measurements are presented, which confirm the electron-hole multiplication originates from within the depletion region (intrinsic section). The main conclusions are summarized in Section \ref{sec:Conclusion}.

\section{Quantum light emitting diode}
\label{sec:QLED}

\begin{figure}
   \begin{center}
   \begin{tabular}{c}
   \includegraphics[height=10cm]{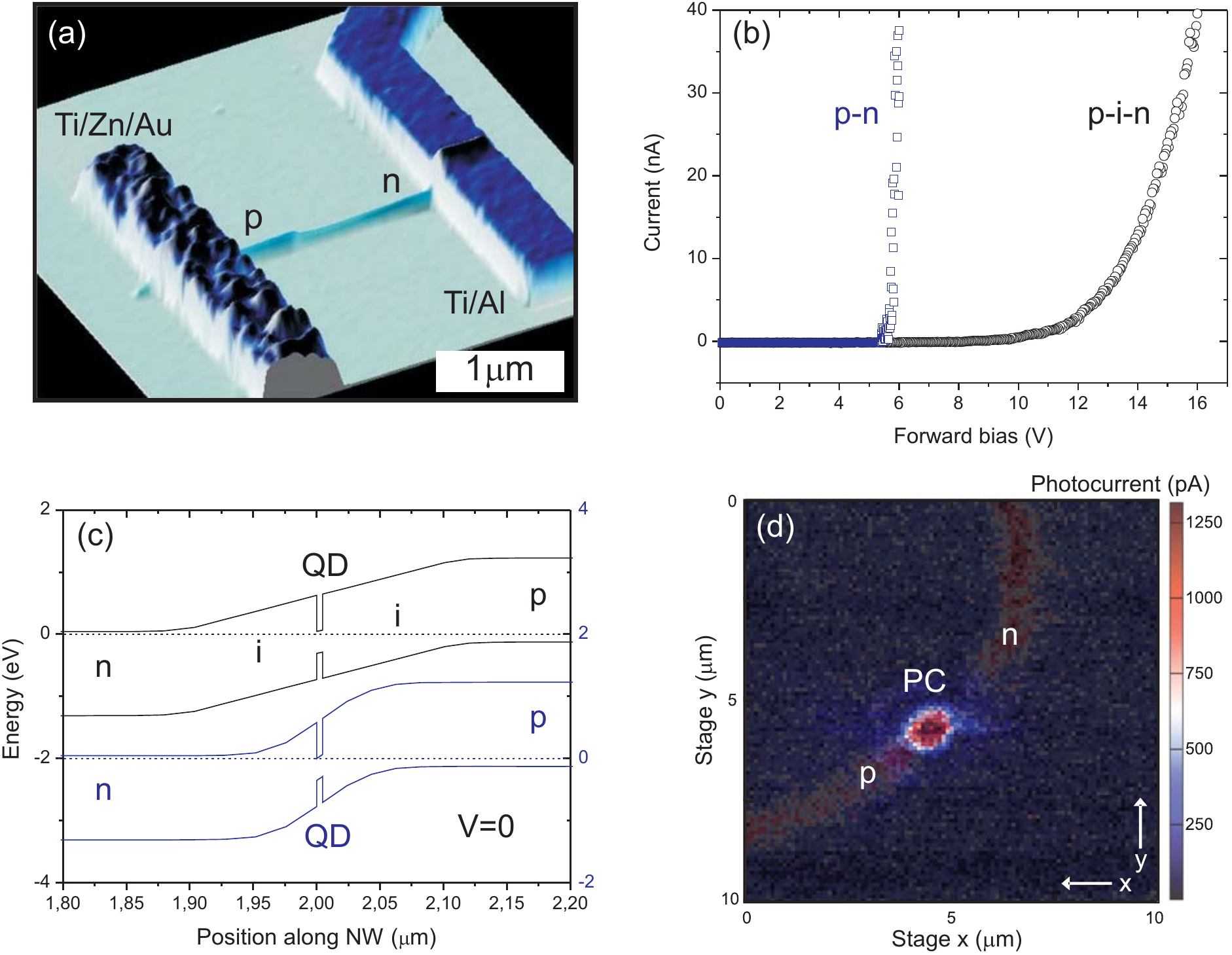}
   \end{tabular}
   \end{center}
   \caption[example]
   { \label{fig:LED}
(a) AFM image of a single contacted nanowire comprised of p- and n-doped sections. The contacts were defined by electron beam lithography and metal evaporation (b) Typical IV characteristic at low temperature (10\,K) for a p-n and p-i-n InP nanowire. (c) Calculated bandstructure of a p-n and p-i-n structure with embedded quantum dot (QD) within the depletion region. The applied bias voltage V is zero. The quasi-Fermi energy level is at 0\,eV (dashed line) (d) Position dependent photocurrent measurement with the p-n junction under a forward bias of 4\,V. The reflection image and photocurrent (PC) image are superimposed onto a single image. Both the reflection image and PC image were acquired simultaneously. The maximum of the PC signal was obtained near the Schottky contact (p-side of nanowire). Laser power = 500\,nW; Excitation wavelength = 532\,nm.}
\end{figure}

In this section, we report recent progress on our QLED comprised of a single InAsP quantum dot embedded within the depletion region of a p-n or p-i-n structure that is synthesized in an InP nanowire. A typical fabricated InP nanowire p-n junction device without the quantum dot is shown in an atomic force microscopy (AFM) image of Fig. \ref{fig:LED}(a). A step-like change of the nanowire is observed from the n-side to p-side of the p-n junction. The InP nanowires were grown at Philips Research Laboratories in Eindhoven by the vapor-liquid-sold (VLS) growth method by use of low pressure metal-organic vapor-phase epitaxy (MOVPE). The n-type dopant gas is hydrogen sulfide and the p-type dopant gas is diethyl-zinc. In samples for which a single quantum dot is incorporated within the depletion region of the p-n junction, the dopant gases are shut off between the n- and p-type InP growth and instead an arsine (AsH$_3$) flux is added to the gas flow. Further details of the growth for these nanowires can be found in Ref.~12. Typically in this study, the nanowire diameter ranges from 30\,nm to 40\,nm and is approximately 4\,$\mu$m in length. After growth, the InP nanowires are transferred onto a SiO$_2$ substrate to be electrically contacted.\cite{Kouwen10}

In order to make ohmic contacts to both sides of the nanowire, we use Ti/Al (100:10\,nm) for the n-side and Ti/Zn/Au (1.5:30:90\,nm) for the p-side. However, the present contacts exhibit a Schottky behavior due to Fermi-level pinning of the p-contact to InP 900\,meV above the valence band.\cite{Brillson81, Minot07} The Schottky nature of the p-contact is evident in the low temperature (10K) current-voltage (IV) characteristic of Figure \ref{fig:LED}(b). These IV curves are typical for p-n and p-i-n InP nanowires. Under forward bias of the p-n junction, an onset of the current is observed at $\sim$6\,V, which is more than four times the bandgap of InP. This Schottky nature of the p-contact is confirmed by position dependent photocurrent measurements in Figure \ref{fig:LED}(d) under forward bias of 4\,V. Here, the maximum photocurrent originates from the p-type Schottky contact. In reverse bias, the nanowire p-n diodes exhibit minimal current ($\sim$10\,pA) until reverse breakdown occurs at $\sim$15\,V.\cite{Minot07} Fig. \ref{fig:LED}(b) also presents the IV characteristic of a p-i-n nanowire device in which a small 200\,nm undoped region of InP has been added in between the p- and n-doped regions. The addition of the intrinsic region shifts the onset voltage to higher bias due to the added resistance of the intrinsic region. A comparison of the band-structure for both the p-n and p-i-n nanowire is shown in Fig. \ref{fig:LED}(c), which was calculated by a one-dimensional Poisson equation using the commercial software NextNano.\cite{Hack02_PhysicaB} The depletion width of the p-n junction is 100\,nm. This 100\,nm is similar to the estimated depletion width by photocurrent measurements in earlier work.\cite{Minot07} In the p-i-n structure, the depletion region is increased to 200\,nm in order to provide a protective barrier between the dopants and the single quantum dot embedded within the depletion region. This protective barrier is intended to improve the quantum dot emission (see Section \ref{sec:p-i-n}). We operate the device by forward biasing the junction. Electrons and holes are injected into the single quantum dot from either side of the nanowire when the quasi-Fermi level is above the lowest energy level state of the quantum dot. Finally, the injected electrons and holes recombine from the ground state of the system to emit a photon with a frequency dependent on the quantum dot size.

\subsection{Optical properties of single quantum dot}
\label{sec:optical-properties}

\begin{figure}
   \begin{center}
   \begin{tabular}{c}
   \includegraphics[height=6cm]{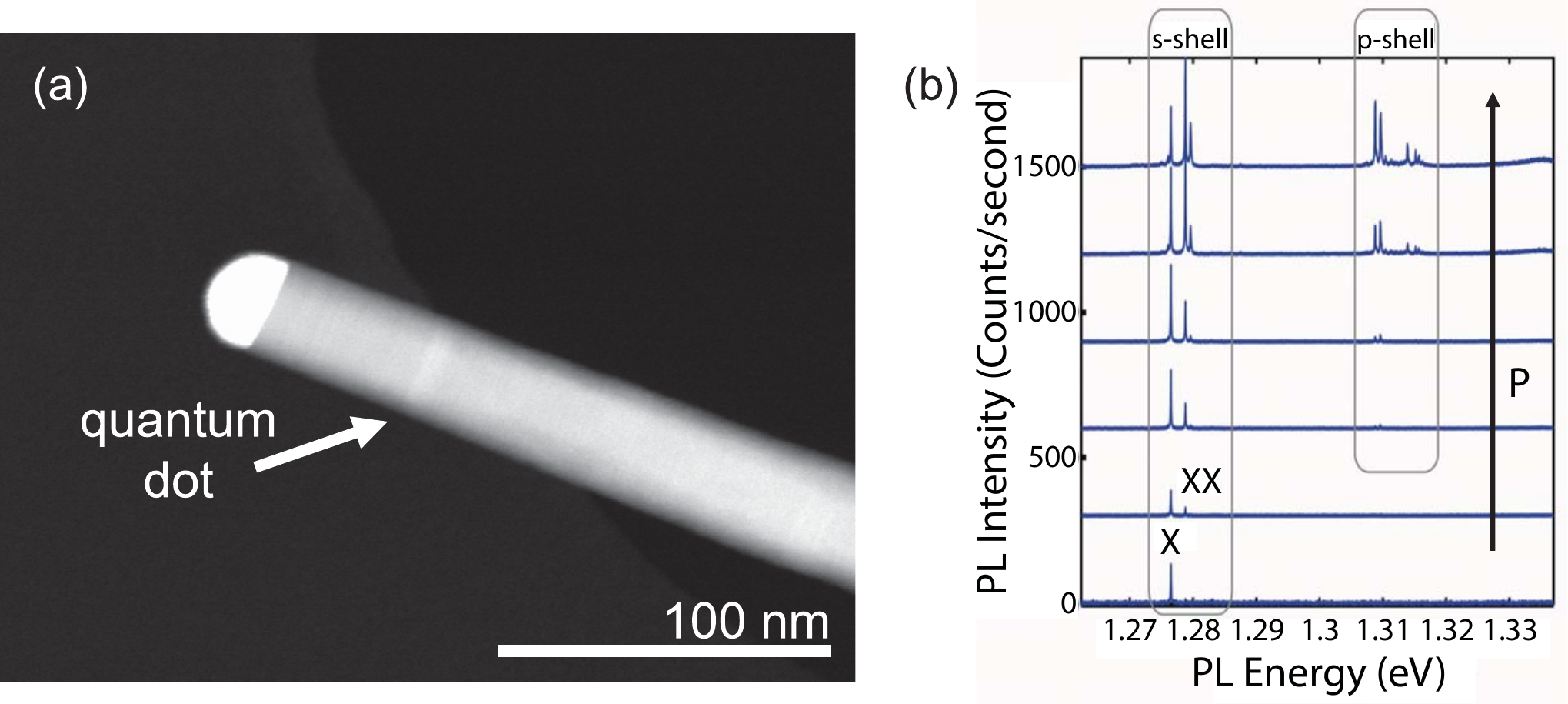}
   \end{tabular}
   \end{center}
   \caption[example]
   { \label{fig:dot}
(a) High resolution TEM image of an InP nanowire with embedded InAsP quantum dot. (b) Power (P) dependant photoluminescence spectroscopy of a typical InAsP/InP quantum dot at low temperature (4.2\,K). The quantum dot was excited with above bandgap excitation (532\,nm). The collected luminescence was dispersed by a single grating spectrometer and detected with a liquid nitrogen Si cooled CCD. Each spectrum is offset along the y-axis for clarity. Power density range corresponds to 0.5 (bottom), 1, 3, 5, 20 and 40 (top) W/cm$^2$.}
\end{figure}

We investigate an InAsP single quantum dot within an undoped InP nanowire. The intrinsic undoped region containing a single quantum dot is intended to be embedded in a p-n junction. In future work, this quantum emitter is to be utilized as a single photon source, which can be driven electrically within the p-n junction defined along the nanowire elongation axis. Fig. \ref{fig:dot}(a) shows a high resolution transmission electron microscope (TEM) image of a single InAsP quantum dot within an InP nanowire. Here, we grow the quantum dot very close to the gold colloid to more easily obtain the TEM images. Usually the quantum dot would be grown in the middle of the nanowire for devices to be processed in the horizontal configuration.\cite{Kouwen10} Typical photoluminescence from a single quantum dot as a function of increasing laser power is shown in Fig. \ref{fig:dot}(b). At low excitation power, only a single emission peak is observed attributed to recombination of a single exciton $X$ involving a single electron-hole pair in the quantum dot $s$-shell. It is this emission peak, when filtered, that produces a single photon source on demand.\cite{Shields07, Claudon10} Of particular significance, only a single quantum dot is probed optically without the requirement of single dot isolation techniques. Increasing the excitation power further modifies the average exciton occupancy of the quantum dot.\cite{Finley01} At approximately 2\,meV higher energy, the biexciton ($XX$) appears in the optical spectrum, which corresponds to recombination of an exciton in the quantum dot $s$-shell in the presence of an additional exciton. This peak can appear either at higher or lower energy than the single $X$ line depending on the quantum dot height.\cite{Rei10_madrid} Finally, $p$-shell emission is observed when the $s$-shell is completely filled. Concomitant with observation of $p$-shell emission are satellite peaks about the $X$ and $XX$ line due to $s$-shell recombination perturbed by the presence of excitons occupying higher lying states in the $p$-shell.\cite{Finley01} More extensive optical properties of dots in wires can be found in recent work by Van Weert \emph{et. al.}\cite{vanWeert09,vanWeert09b}, and Van Kouwen \emph{et al.}\cite{Kouwen10}
\subsection{Single quantum dot in a p-n junction}
\label{sec:p-n}

\begin{figure}
   \begin{center}
   \begin{tabular}{c}
   \includegraphics[height=10cm]{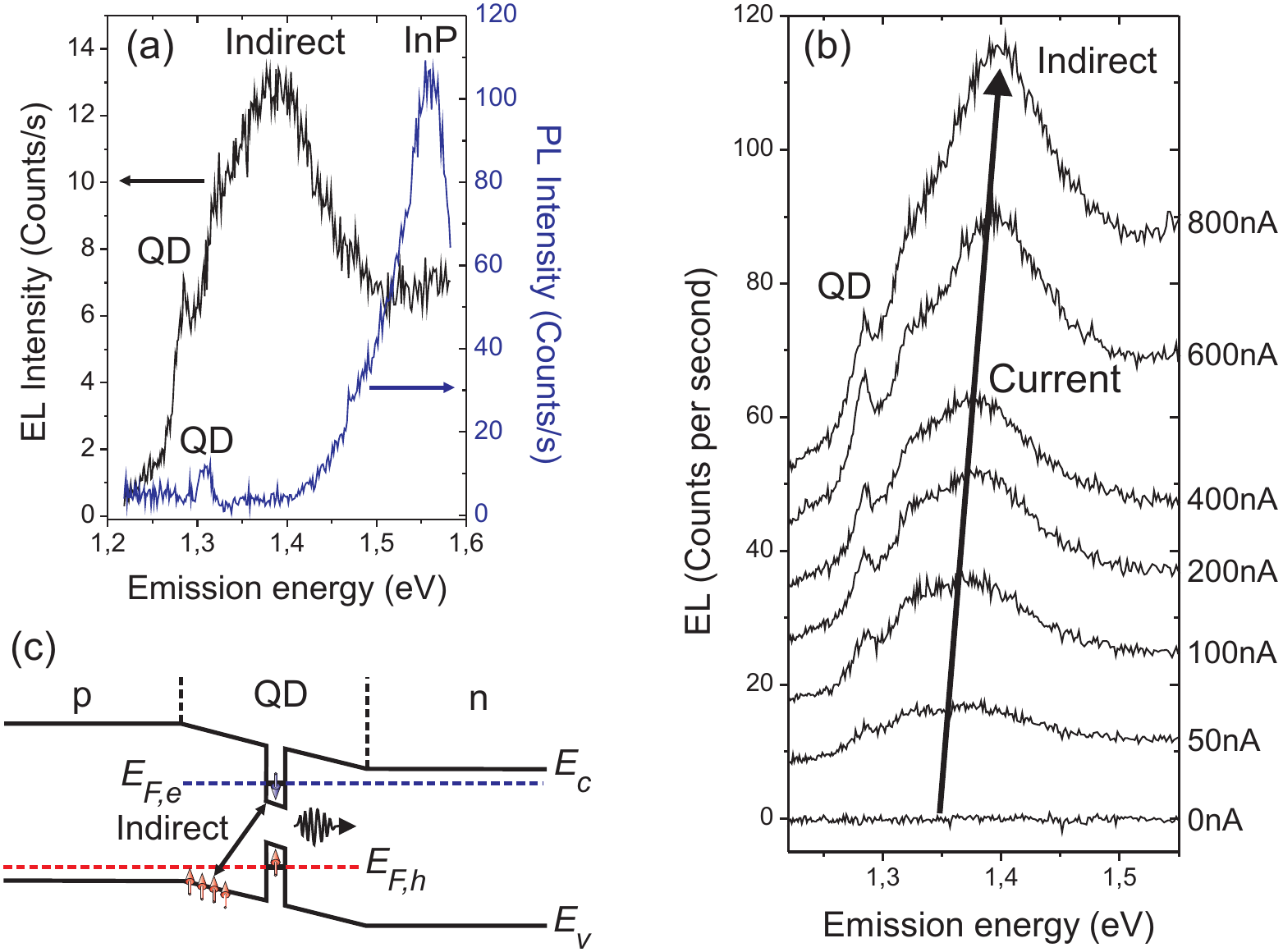}
   \end{tabular}
   \end{center}
   \caption[example]
   { \label{fig:EL_pn}
(a) Low temperature (10\,K) electroluminescence (EL) and photoluminescence (PL) spectrum from a single quantum dot (QD) embedded in a p-n junction. EL is taken under forward bias of 6.2\,V and PL is taken at V=0. Injection current in EL is 100\,nA and excitation power in PL is 100\,nW (excitation wavelength = 532\,nm). (b) Low temperature (10\,K) EL as a function of increasing current injected by the p-n junction under forward bias. The EL spectra are offset along the y-axis for clarity. (c) Energy band diagram, which illustrates the indirect recombination process involving recombination from electrons in the quantum dot with holes in the p-doped region. Indirect transitions between electrons in the n-doped region and holes in the QD are equally probable. E$_c$, conduction band energy; E$_v$, valence band energy.}
\end{figure}

We investigate an InAsP single quantum dot embedded within a p-n junction defined along the nanowire elongation axis. Electroluminescence (EL) from a single quantum dot
is detected when the p-n junction is forward biased above the onset
voltage. Fig. \ref{fig:EL_pn}(a) compares a typical EL spectrum with a photoluminescence (PL) spectrum from the same
quantum dot. In contrast to the PL emission, the quantum dot EL
emission peak is observed on top of a broad emission spectrum. The
broad emission peak at higher energies is likely due to indirect
transitions involving recombination from the quantum dot with either
the n- or p-doped regions (see Fig. \ref{fig:EL_pn}(c)). The QD emission is also broadened as compared to the expected PL presented in the previous section. The
increased linewidth is likely caused by interactions with charges
that are in the vicinity of the dot either from the doping in the nanowire or charge traps in the SiO$_2$ substrate.\cite{Kouwen10}

Our interpretation of indirect transitions is corroborated in Fig. \ref{fig:EL_pn}(b) where we present EL as a function of increasing current injected by the p-n junction under forward bias. Here, we observe a blue shift for increasing current, which is consistent with a lowering of the valence band on the p-side with respect to the quantum dot ground state. This blue shift is clearly not observed for the quantum dot emission peak. Another possible mechanism for the observed blue shift for increasing current is attributed to state filling at higher injection currents. Such behaviour is analogous to increasing the laser power presented in the previous section. However, there is an increased broadening of the EL emission here since there are more states available in the p- and n-doped regions as compared to the discrete energy levels of the quantum dot.

From the EL measurements presented here, we estimate the quantum efficiency of electrons injected to photons generated to be about 10$^{-3}$. We assume in this calculation a light detection efficiency for the EL measurement of $\sim$0.1\,\%. The estimated efficiency is much less than would be expected from a one-dimensional system where 100\,$\%$ of the injected current contributes to emitted photons. Possible reasons for inefficiency are: minority carrier diffusion escape to the metal contacts due to an imbalance for the number of electrons and holes; and recombination via indirect transitions.

\subsection{Single quantum dot in a p-i-n junction}
\label{sec:p-i-n}
To improve the EL emission, we synthesize 100\,nm of intrinsic InP on either side of the InAsP quantum dot, which is embedded within the p-n junction. The energy band diagram of this p-i-n structure was shown previously in Fig. \ref{fig:LED}(c). Here, the depletion region of the p-i-n structure is 200\,nm as compared to 100\,nm for the p-n structure. The increased separation between the quantum dot and doped sections of the InP nanowire is expected to improve the dot quality and therefore quantum efficiency of the quantum LED.

\begin{figure}
   \begin{center}
   \begin{tabular}{c}
   \includegraphics[height=8cm]{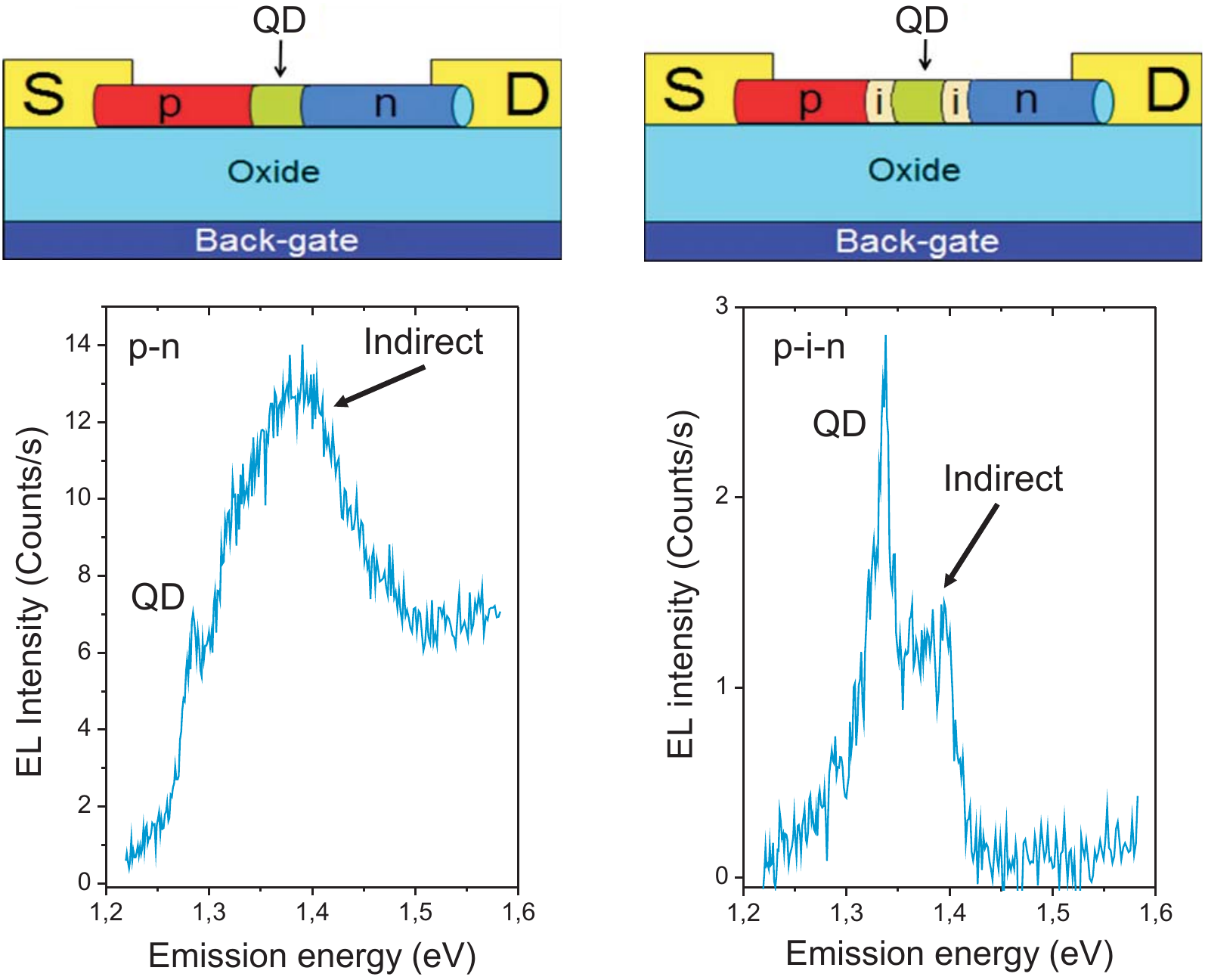}
   \end{tabular}
   \end{center}
   \caption[example]
   {\label{fig:EL_pin}
Comparison of low temperature (10\,K) EL emission for p-n (left panel) and p-i-n (right panel) with an embedded InAsP quantum dot within the depletion region. The injected current for both EL spectra is $\sim$100\,nA. The broad emission peak (indirect transitions) are suppressed in the p-i-n structure as compared to the p-n structure. A schematic view of the p-n and p-i-n structures is shown in the upper panel. S, source for the p-contact; D, drain for the n-contact.}
\end{figure}

A comparison of the EL emission for both the p-n and p-i-n structure is shown in Fig. \ref{fig:EL_pin}. In the case of the p-i-n structure (right panel), a suppression of the broad emission peak at higher energy is observed as compared to the p-n structure (left panel). This suppression is due to an additional $\sim$100\,nm of undoped InP between the quantum dot and the doped sections of the nanowire, which reduces the probability for recombination via indirect transitions and direct transitions involving the p- and n-doped regions. Although we expect an increased quantum efficiency for the p-i-n structure, we observe similar count rates for the quantum dot emission from both structures. Further work to improve the count rate of these single quantum dots is required in order to obtain a more accurate assessment of quantum efficiency.

\section{Si nanowire avalanche photodiode}
\label{sec:SiNW}

In this section, we report recent progress toward single photon
detection with only a single Si nanowire comprised of p-doped,
intrinsic, and n-doped sections (p-i-n diode). Under the appropriate
growth conditions of the Si nanowire, an avalanche breakdown mechanism can be achieved under large reverse bias of the p-i-n diode.\cite{Yang06, Hayden06} The avalanche mechanism is a
key requirement to detect single photons from a single nanowire. We show here that we can fabricate Si p-i-n avalanche photodiodes, which exhibit large multiplication factors,
in excess of 1000, near the breakdown voltage of the p-i-n diode.
We perform position dependent photocurrent measurements that verify
the electron-hole multiplication occurs within the depletion region of the Si
nanowire with spatial resolution of 600\,nm (laser spot diameter).

\subsection{Device}
\label{sec:IV}

\begin{figure}
   \begin{center}
   \begin{tabular}{c}
   \includegraphics[height=8cm]{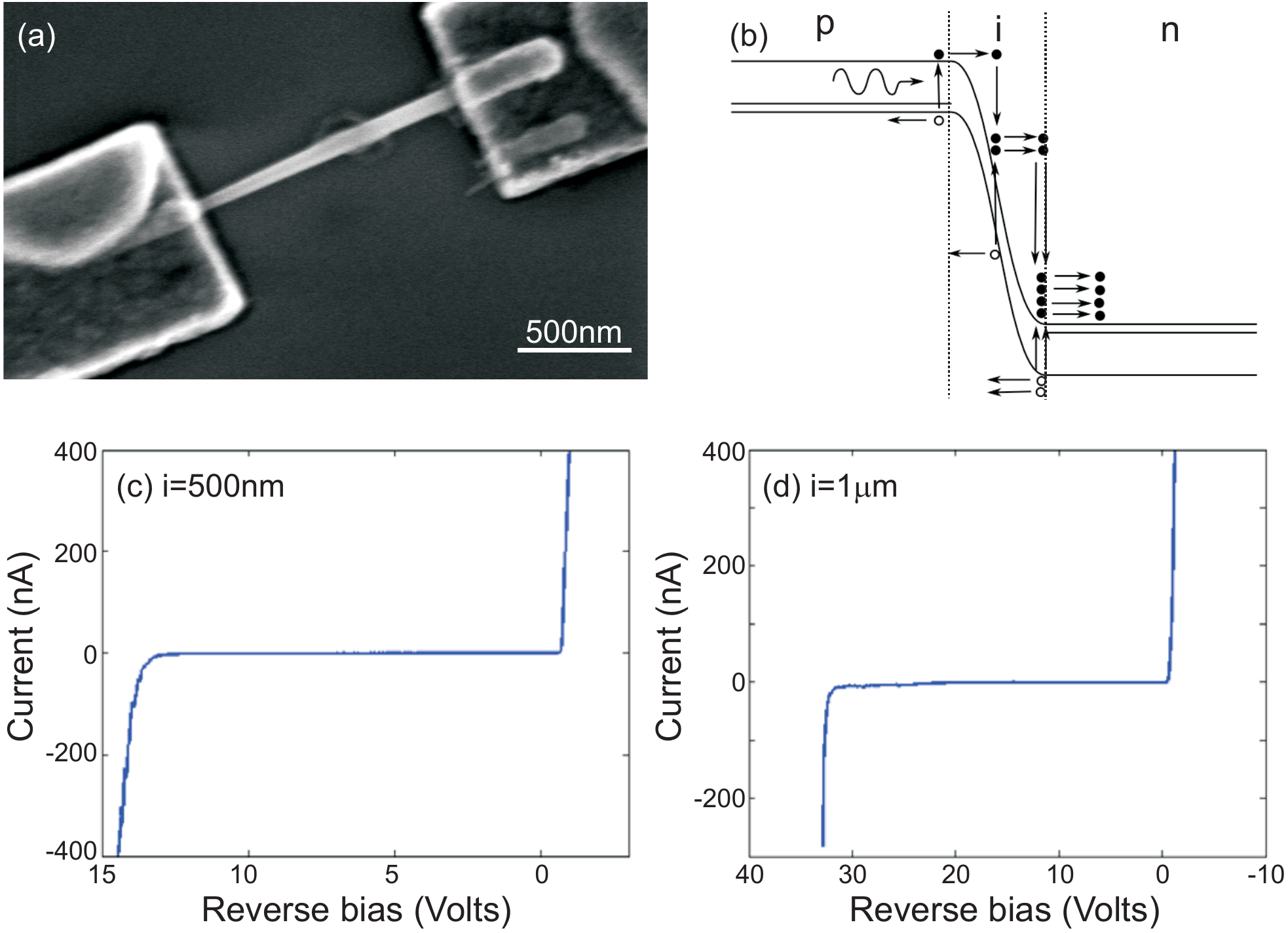}
   \end{tabular}
   \end{center}
   \caption[example]
   { \label{fig:SiIV}
(a) SEM of Si nanowire avalanche photodiode device. Both ends of the nanowire are contacted with Ti/Au (5:100\,nm) (b) Schematic view of the bandstructure for the p-i-n nanowire. One single incident photon on the nanowire that is absorbed, creates an electron-hole pair and induces an avalanche effect close to the reverse diode breakdown of the p-i-n diode.  Room temperature IV characteristic for a single Si nanowire with an intrinsic region, i, of (c) $i=500$\,nm and (d) $i=1$\,$\mu$m. The doping of these Si nanowires are $N_A=1\times10^{18}$\,cm$^{-3}$ and $N_D=1\times10^{19}$\,cm$^{-3}$. The background doping in the intrinsic region is estimated to be in the order of $1\times10^{16}$\,cm$^{-3}$.}
\end{figure}

Shown in Fig. \ref{fig:SiIV}(a) is a typical scanning electron
micrograph (SEM) of a single contacted Si nanowire, which was
transferred onto a SiO$_2$ substrate. Both ends of the nanowire are
contacted with Ti/Au (5\,nm:100\,nm) defined by electron beam
lithography and evaporation. The Si nanowires were grown at IBM
Z\"urich by the vapour-liquid-solid (VLS) growth mechanism using
chemical vapour deposition (CVD). The Si nanowire growth was
achieved by using a diluted precursor gas (silane: SiH$_4$). The
p-type Si segment of the nanowire is synthesized first, after which
an intrinsic Si region and n-type Si segment is grown. The p-side of
the nanowire is $\sim$ 60\,nm in diameter, whereas the n-side is
$\sim$ 20\,nm in diameter. To facilitate the p- and n-type doping,
diborane was used as the precursor gas for the p-dopant and
phosphine was used as the precursor gas for the n-dopant.\cite{Schmid08}

Operation of a Si nanowire avalanche photodiode is achieved close to
the reverse breakdown voltage of the p-i-n diode. A schematic view of the
energy band diagram for the p-i-n diode is shown in Fig.
\ref{fig:SiIV}(b) under a large reverse bias. Here, a single
incident photon that is absorbed in the intrinsic region or within the diffusion length of the intrinsic region generates
an electron-hole pair, which is then separated and accelerated across the depletion region by the large electric field. After the charge carrier gains
enough kinetic energy to overcome the impact ionization threshold, and upon collision with the lattice, an additional electron-hole pair is created by impact ionization. As the impact ionization process is repeated,
the result is a multiplication of electrons and holes and hence, a
current may be detected by an external electric circuit from
only one incident single photon. To measure a large enough signal from only a single photon, the diode is operated beyond the breakdown voltage of the p-i-n diode in the Geiger mode. In the work presented here, however, we operate the Si nanowire avalanche photodiode close to the reverse breakdown voltage in order to investigate the photo-response from more than one photon and quantify the multiplication factor of a single Si nanowire.

Two different Si nanowire p-i-n structures were fabricated with the aim of reaching the avalanche  breakdown regime under large reverse bias. In Fig. \ref{fig:SiIV}(c) and Fig. \ref{fig:SiIV}(d), we present the room temperature current-voltage (IV) characteristic for a p-i-n structure with an intrinsic region of 500\,nm and 1\,$\mu$m, respectively. In both structures, a diode behaviour is observed, which is typical of our Si nanowire p-i-n structures. The reverse bias breakdown voltage is different for the two structures. The Si nanowire breakdown voltage with the 500\,nm intrinsic region was -14\,V ($E=280$\,kV/cm), while the Si nanowire with the 1\,$\mu$m intrinsic region exhibited a reverse breakdown voltage of -32\,V ($E=320$\,kV/cm). Although the reverse breakdown voltages significantly differ, the electric fields at reverse breakdown are similar. The electric field was calculated assuming all of the voltage dropped across the intrinsic region of the nanowire. Of particular significance is that an avalanche breakdown is achieved in both devices under large reverse bias. We confirmed this observed IV behaviour was attributed to an avalanche breakdown mechanism by temperature dependent IV measurements similar to O. Hayden \emph{et al.}\cite{Hayden06}

\subsection{Multiplication}
\label{sec:Multiplication}

\begin{figure}
   \begin{center}
   \begin{tabular}{c}
   \includegraphics[height=5cm]{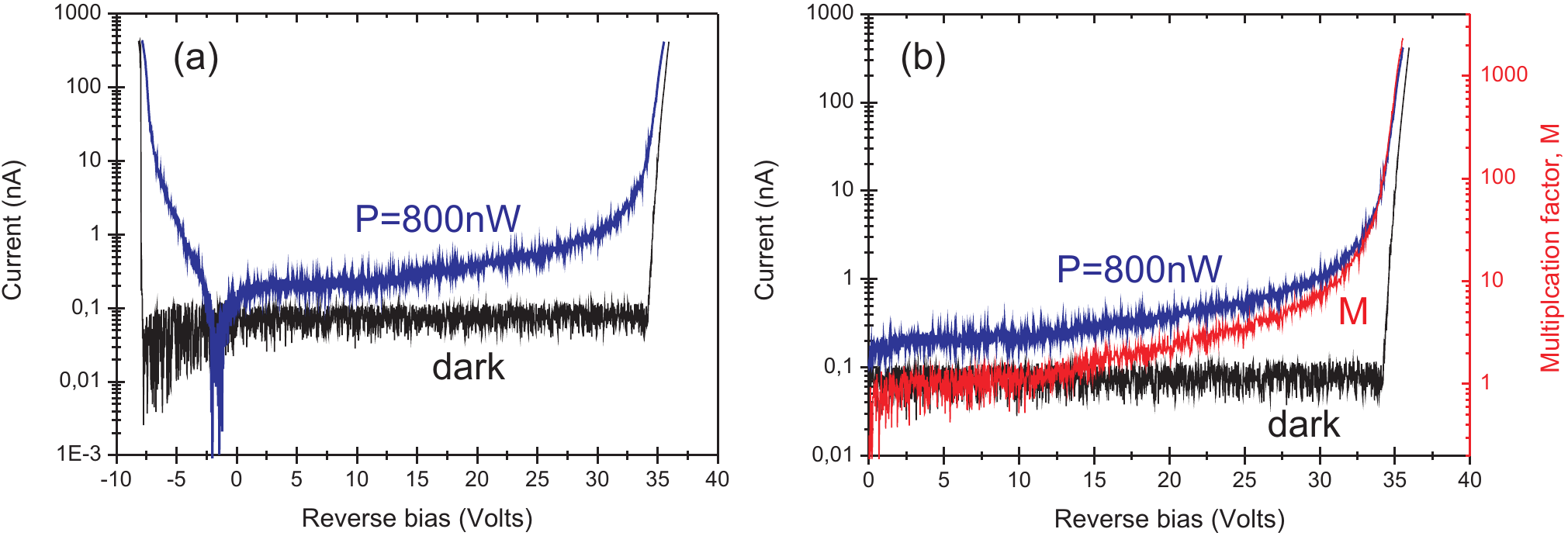}
   \end{tabular}
   \end{center}
   \caption[example]
   { \label{fig:multiplication}
(a) Low temperature (T = 20\,K) photo-response of Si nanowire avalanche photodiode with intrinsic region length of 1\,$\mu$m and power, P, of 800nW (power density = 280\,W/cm$^2$). The dark current is also shown in order to estimate the multiplication factor, M. The excitation wavelength is 532\,nm. (b) Multiplication factor extracted from (a) using equation (1). A multiplication factor of 2300 is measured close to the reverse breakdown voltage of 35\,V.}
\end{figure}

To quantify the multiplication factor of the Si nanowires that exhibit an avalanche process near reverse breakdown, we characterize the photo-response of a single Si nanowire as a function of reverse bias at low temperature (T = 20\,K). We measure with relatively low laser power (800\,nW $\sim$ 280\,W/cm$^2$) and compare this to the measured dark current. The result of this measurement is shown in Fig. \ref{fig:multiplication}(a). At low reverse biases (from 0\,V to 10\,V), we observe a plateau for the measured photocurrent. This regime corresponds to no multiplication of carriers since the electric field across the depletion region is too small to induce impact ionization. Here, the small electric field separates the introduced charge carriers, which generates the measured photocurrent. Above 10\,V, an increase in the photocurrent is observed until the reverse breakdown voltage of the diode is reached and an exponential increase of the photocurrent can be observed. The increase in observed photocurrent is attributed to a multiplication of carriers as the absorbed photons initiate the impact ionization process as they accelerate across the depletion region due to the large electric field. Close to the reverse breakdown voltage, the exponential increase of the measured photocurrent is a direct result of the avalanche multiplication process.

The multiplication factor, M, is defined as $M=N_{eh}/N_{ph}$, where $N_{eh}$ is the number of extracted electrons and holes and $N_{ph}$ is the number of incident photons that are absorbed. At low reverse biases across the p-i-n diode, there is no multiplication of carriers as mentioned previously. This regime corresponds to unity gain ($M=1$) in which $N_{eh}=N_{ph}$ if we assume 100$\%$ of the incident photons are absorbed and there is no loss in charge transport of carriers. To include the absorption loss of the photon, we can relate $N_{eh}$ and $N_{ph}$ to the measured photocurrent. At low reverse biases, the measured photocurrent with unity gain, $I_0$, is proportional to $N_{ph}$ since the photons that are absorbed do not have enough energy to induce impact ionization and is therefore swept away by the depletion region and contributes to the measured photocurrent. In contrast, at large reverse biases, the absorbed photons do have enough energy to induce impact ionization and initiate the avalanche process. Here, the measured photocurrent, $I_{ph}$, is proportional to $N_{eh}N_{ph}$. Accounting for the dark current and assuming the incident photon rate and absorption of the intrinsic region in the nanowire is constant as a function of reverse bias, the multiplication factor can be approximated by
\begin{equation}
    M={I_{ph}-I_{dark}\over{I_0-I_{dark(0)}}},
    \label{eq:M}
\end{equation}
\noindent
where $I_{dark(0)}$ is the average measured dark current in the voltage range where $M=1$, and $I_{dark}$ is the measured dark current as a function of reverse bias.

The multiplication factor has been extracted from the data of Fig. \ref{fig:multiplication}(a) using equation (1) and plotted in Fig. \ref{fig:multiplication}(b) as a function of reverse bias. As mentioned previously, no multiplication is observed at low reverse biases (from 0\,V to 10\,V). Quite remarkably, close to the breakdown voltage under large reverse bias ($\sim$35\,V), we obtain a multiplication factor of 2300 from only a single Si nanowire. This multiplication factor that we measure is considered to be very high as compared to previously reported M factors from a single Si nanowire of 30.\cite{Yang06}. To reach the single photon detection regime, the challenge remains to operate the device beyond the reverse breakdown voltage in order to achieve multiplication factors large enough to produce an electrical signal that is detectable from a single incident photon.

\begin{figure}
   \begin{center}
   \begin{tabular}{c}
   \includegraphics[height=11cm]{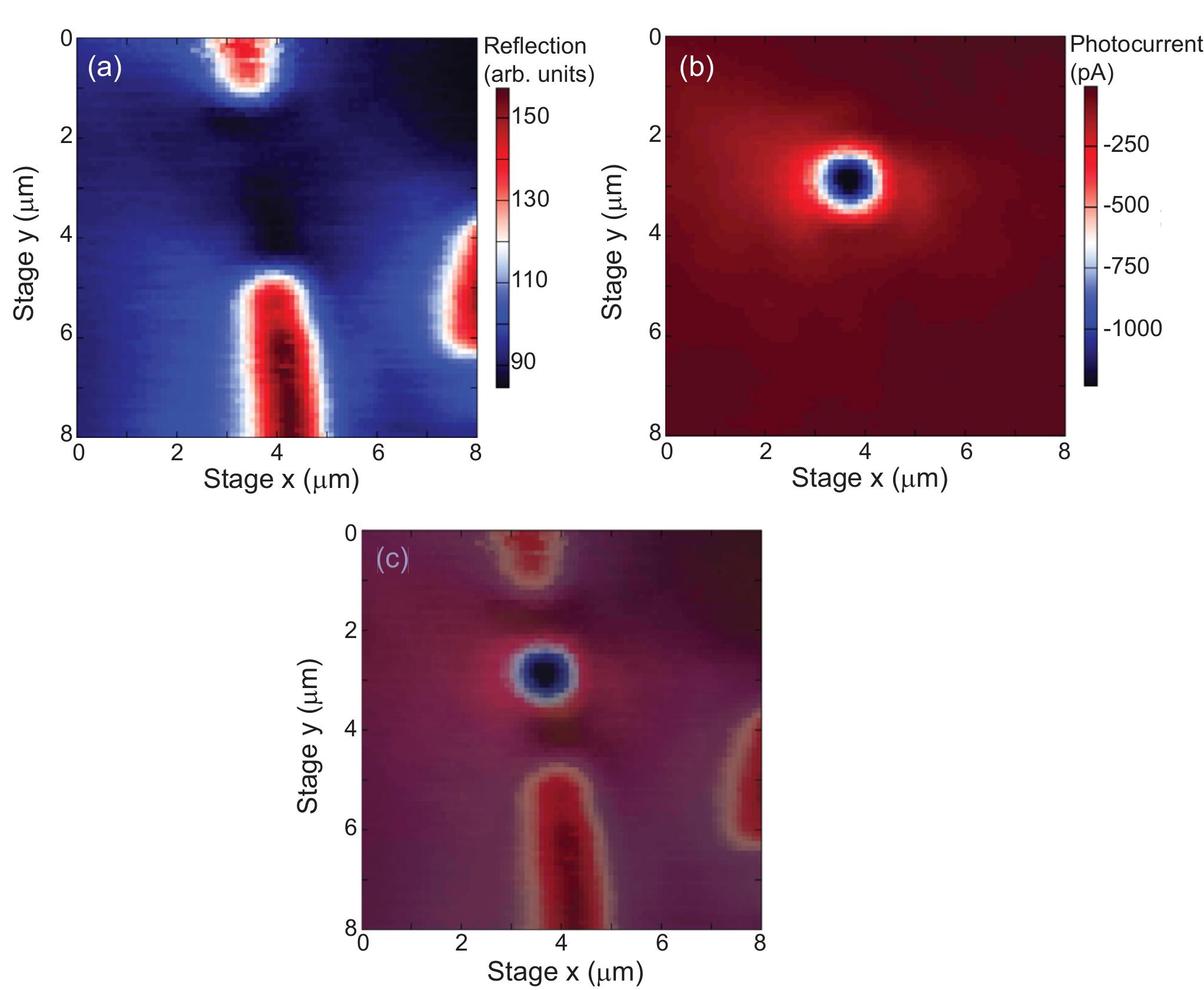}
   \end{tabular}
   \end{center}
   \caption[example]
   { \label{fig:PC}
(a) Reflection image and (b) photocurrent image, which are both simultaneously acquired as a function of laser position at low temperature (T = 20\,K). (c) The reflection and photocurrent images from (a) and (b) are superimposed on the same image. We conclude from this image that the photocurrent, and thus multiplication, originate from within the depletion region (intrinsic section) of the Si nanowire.}
\end{figure}
\subsection{Position dependent photocurrent}
\label{sec:PC}

To confirm the electron-hole multiplication that we observe originates from the intrinsic region of the nanowire and not from Schottky contacts, we perform position dependent photocurrent measurements under large reverse bias while simultaneously recording the reflection image of the sample by a commercially available photodiode. The reflection image is shown in Fig. \ref{fig:PC}(a) and the photocurrent image is shown in Fig. \ref{fig:PC}(b) for a reverse bias voltage of -32\,V and excitation power of 750\,nW (excitation wavelength is 532\,nm). By superimposing the two images, we can conclude that the measured photocurrent, and thus electron-hole multiplication, originates from within the depletion region (intrinsic section) of the Si nanowire.

\section{Conclusion}
\label{sec:Conclusion}

We have presented an overview of single nanowire devices for use in light emission and detection applications toward the single photon level. We studied the optical properties of single quantum dots embedded in an InP nanowire comprised of p-n and p-i-n structures. We demonstrated that under forward bias of the QLED, light is emitted from the quantum dot. A broad emission peak was observed at higher energies than the quantum dot emission, which significantly reduces the quantum efficiency of the QLED. The dot properties were shown to improve when surrounding the dot material by a small 200\,nm intrinsic region of InP. Further improvements to the out-coupling efficiency and dot quality is required in order to reach the single photon regime.

We also presented Si nanowire avalanche photodiodes comprised of p-doped, intrinsic, and n-doped sections. Under large reverse bias close to the breakdown voltage, we observed multiplication factors in excess of 1000. This multiplication factor is two orders of magnitude larger than previously reported Si nanowires.\cite{Yang06} Position dependent photocurrent measurement were performed that confirm the measured electron-hole multiplication originates from within the depletion region (intrinsic section). The multiplication factors that we demonstrate are very promising to reach the single photon regime in future experiments from only a single Si nanowire.

\acknowledgments     

We acknowledge E. D. Minot for the AFM image of the single contacted nanowire and R. N. Schouten for technical assistance. This work was
supported by the Dutch ministry of economic affairs (NanoNed
DOE7013), the Dutch Organization for Fundamental Research
on Matter (FOM), the European FP6 NODE (015783)
project, the DARPA QUEST grant HR0011-09-1-0007 and
The Netherlands Organization for Scientific Research (NWO).
The work of REA was carried out under project number
MC3.0524 in the framework of the strategic research program
of the Materials Innovation Institute (M2I), www.m2i.nl.


\bibliography{report}   
\bibliographystyle{spiebib}   

\end{document}